\documentclass[twocolumn,prl,preprintnumbers,amsmath,amssymb,floatfix]{revtex4}

\usepackage[linktocpage,bookmarksopen,bookmarksnumbered]{hyperref}
\usepackage{graphicx}
\usepackage{dcolumn}

\usepackage{amsmath,graphics,epsfig,color,verbatim,ulem}

\begin{document}

\setlength{\pdfpageheight}{\paperheight}
\setlength{\pdfpagewidth}{\paperwidth}

\title{Kinetic frustration and the nature of the magnetic and paramagnetic states in iron pnictides and iron chalcogenides}
\author{Z. P. Yin$^{1,2}$}
\email{yinzping@physics.rutgers.edu}
\author{K. Haule$^1$}
\author{G. Kotliar$^1$}
\affiliation{$^1$Department of Physics and Astronomy, Rutgers University, Piscataway, New Jersey 08854, United States.}
\affiliation{$^2$Department of Physics and Astronomy, Stony Brook University, Stony Brook, New York 11794, United States.}
\date{\today}
\maketitle

\textbf{
The iron pnictide and chalcogenide compounds are a subject of intensive investigations due to 
their high temperature superconductivity.\cite{a-LaFeAsO} 
They all share the same structure, but there is significant variation in their physical
properties, such as magnetic ordered moments, effective masses, superconducting gaps and T$_c$.
Many theoretical techniques have been applied to individual compounds but no consistent
description of the trends is available~\cite{np-review}.
We carry out a comparative theoretical study of a large number of iron-based compounds 
in both their magnetic and paramagnetic states. 
We show that the nature of both states is well described by our method 
and the trends in all the calculated physical properties such as the 
ordered moments, effective masses and Fermi surfaces are in good
agreement with experiments across the compounds.
The variation of these properties can be traced to
variations in the key structural parameters, rather than changes in
the screening of the Coulomb interactions. 
Our results provide a natural explanation 
of the strongly Fermi surface dependent superconducting gaps observed in experiments\cite{Ding}. 
We propose a specific optimization of the crystal structure to look for higher T$_c$ superconductors.
}

The iron pnictides are Hund's metals~\cite{Haule-njp}, where the
interaction between the electrons is not strong enough to fully
localize them, but it significantly slows them down, so that
the low energy quasiparticles have much enhanced mass.  These
quasiparticles are composites of charge and a fluctuating magnetic
moment originating in the Hund's rule interactions which tend to align
electrons with the same spin and different orbital quantum numbers
when they find themselves on the same iron atom.

\begin{figure}[bht]
\includegraphics[width=0.95\linewidth]{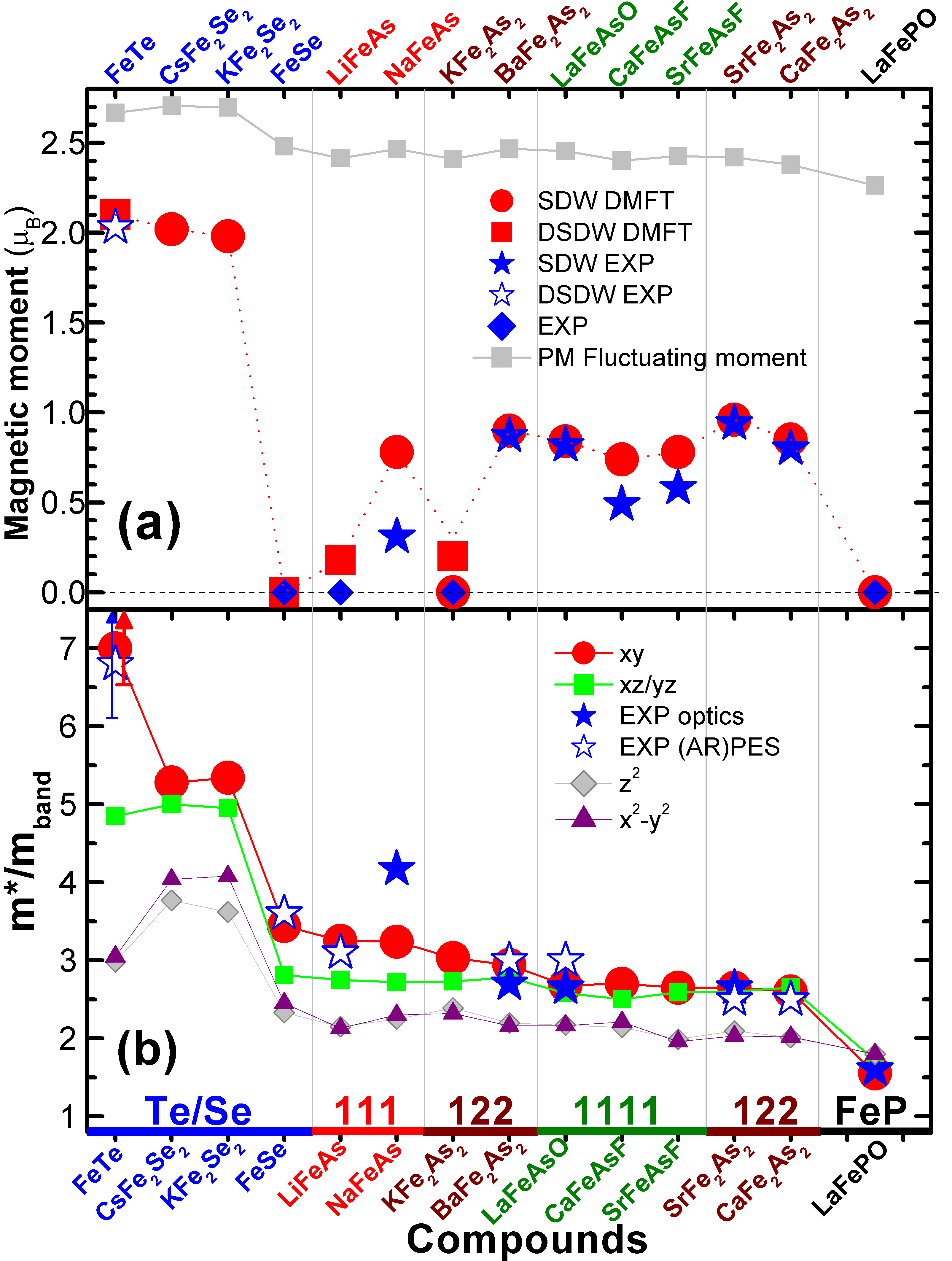}
\caption{
\textbf{Ordered magnetic moments and mass enhancements in iron-based compounds.}
(a)the DFT+DMFT calculated and experimental\cite{moment-FeTe,
moment-NaFeAs,
moment-Ba122,
moment-LaFeAsO,
moment-CaFeAsF, moment-SrFeAsF,
moment-Sr122,
moment-Ca122}
Fe magnetic moments in the SDW and DSDW states.
Also shown is the calculated fluctuating moment in the PM state. 
(b)The DFT+DMFT calculated mass enhancement m*/m$_{band}$ of the Fe $3d$ orbitals in the PM state
and the low energy effective mass enhancement obtained
from optical spectroscopy experiments\cite{optics-NaFeAs, optics-Ba122-Sr122, optics-LaFeAsO, optics-LaFePO} 
and (angle-resolved) photoemission spectroscopy
experiments\cite{PES-FeTe, PES-FeSe, PES-LiFeAs, PES-Ba122-Sr122, PES-Ba122, PES-Ca122}. 
}
\label{mass-and-moment}
\end{figure}

A central puzzle in this field is posed by the variation of the
ordered magnetic moment across the iron pnictides/chalcogenides
series.  In the fully localized picture the atom resides in a single
valence, therefore the ordered moment is equal to the atomic moment
(4$\mu_B$ per iron), possibly reduced by quantum fluctuations. This
picture is realized in cuprate superconductors where quantum
fluctuations reduce the Cu$^{2+}$ moment  by 20\%.  In the fully
itinerant weak coupling picture, such as spin density wave (SDW) in chromium metal, the ordered
moment is related to the degree of Fermi surface nesting.  It is by
now clear that the iron pnictides are not well described by either
fully localized or fully itinerant picture, nor by the density
functional theory (DFT), which greatly overestimates the ordered
magnetic moments.  It has been advocated that the shortcomings of 
DFT can be circumvented by incorporating
the physics of long wavelength fluctuations~\cite{Mazin-NP}. Here we
take the opposite perspective.  While critical long-wavelength
fluctuations certainly play a role near the phase transition lines, we
will show that the local fluctuations on the iron atom can account for
the correct trend of magnetic moments and correlation strength in iron
pnictides/chalcogenide layered compounds.

Using the combination of density
functional theory and dynamical mean field theory (DFT+DMFT) (see online material for details), we
studied two different real space orderings, the SDW ordering, characterized by wave vector $(\pi,0,\pi)$ (this
vector is written in coordinates with one Fe atom per unit cell),
which is experimentally found in iron arsenide compounds, and
$(\pi/2,\pi/2,\pi)$ ordering, denoted by double stripe spin density
wave (DSDW). The latter was found experimentally in FeTe.
Figure~\ref{mass-and-moment}(a) shows our theoretical results for the
ordered moment in both phases together with experimentally determined
values \cite{moment-FeTe,
moment-NaFeAs,
moment-Ba122,
moment-LaFeAsO,
moment-CaFeAsF, moment-SrFeAsF,
moment-Sr122,
moment-Ca122}
from across all known families of iron-based superconducting
compounds.
There is an overall good agreement between theory and experiment, in
particular LaFePO is predicted to be nonmagnetic, the majority of 1111
and 122 compounds have ordered moment below 1.0 $\mu_B$, and FeTe orders
with DSDW moment of 2.1 $\mu_B$.

\begin{figure}[htb]
\includegraphics[width=0.95\linewidth]{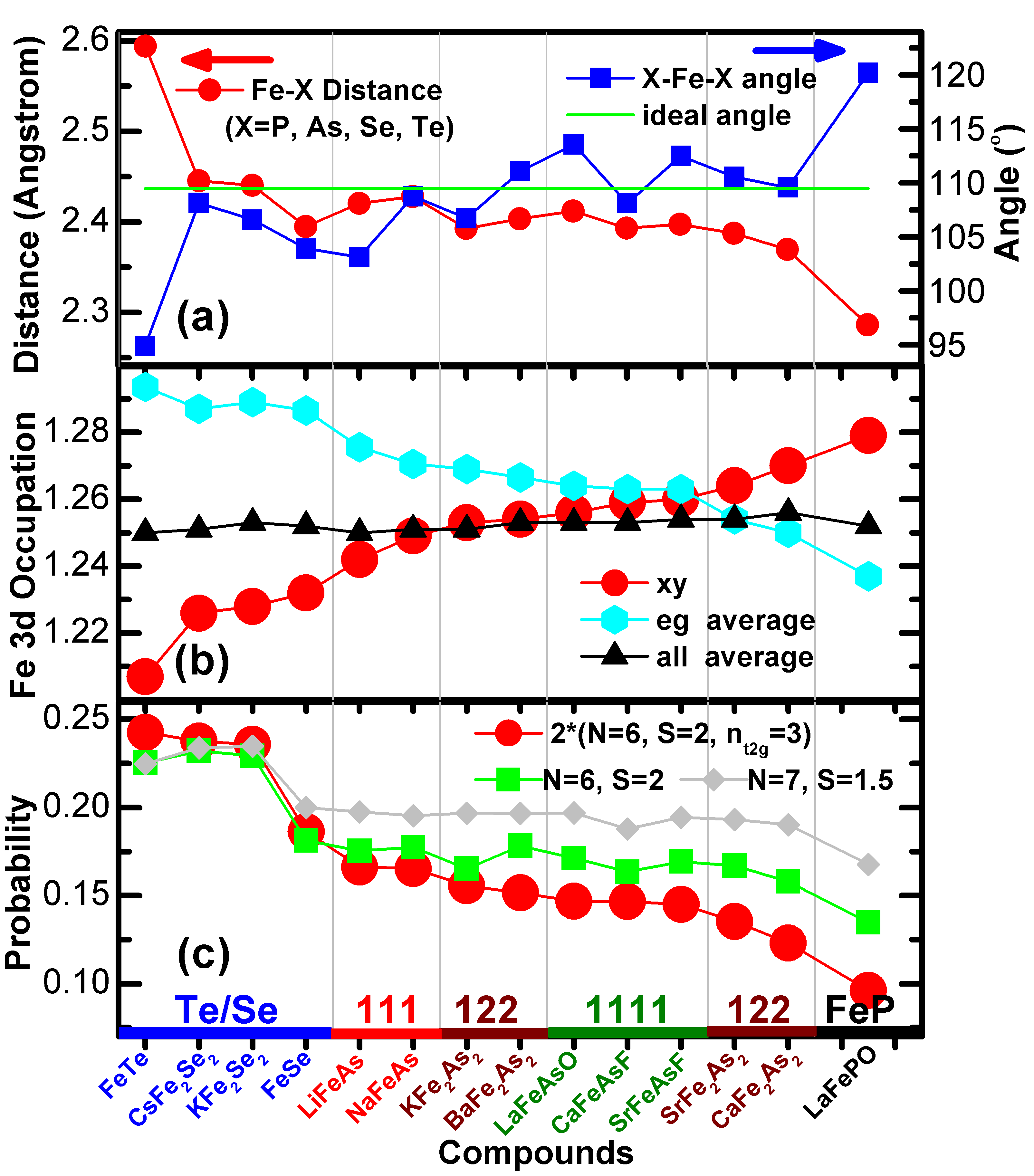}
\caption{
\textbf{Structure, orbital occupation and probability of selected atomic states of Fe.}
(a)The Fe-$X$ ($X$=P, As, Se and Te) distance and $X$-Fe-$X$ angle in iron-based compounds, where the two $X$ atoms are in the same $ab$ plane.
Note this angle is different from the $X$-Fe-$X$ angle where the two $X$ atoms are in different $ab$ planes;
(b)The orbital occupation of the $xy$ orbital and the average values for the $eg$ orbitals and all five orbitals;
(c)The probability of selected atomic configurations of Fe where N (S) is the total number (spin) of Fe $3d$ electrons in the atomic configuration.
}
\label{nd-and-prob}
\end{figure}

We now explain the variation of the
ordered moment in terms of real space and momentum space concepts.
The size of the fluctuating local moment, which can be extracted from
neutron scattering experiments, gives an upper bound to the size of the
ordered magnetic moment and is also plotted in
Fig.~\ref{mass-and-moment}(a).  It was computed from the atomic
histogram displayed in Fig.\ref{nd-and-prob}(c), which shows the
percentage of time the iron $3d$ electrons spend in various atomic
configurations when the system is still in its paramagnetic (PM)
state. Only high spin states, which carry a large weight as a result
of the Hund's rule coupling in iron, are displayed (see also online material for complete histogram).
A monovalent histogram with only the atomic ground state would give iron
magnetic moment of 4~$\mu_B$. 

In a correlated Fermi
liquid, the spin excitations are described in terms of individual
particle hole pair excitations and their collective motion.  Their
residual interaction can lead to a magnetic state when the particle
hole excitations condense at non-zero wave vector.  A large
quasiparticle mass, naturally facilitates this condensation, hence we
expect that the size of the ordered moment will correlate with the
mass of the quasiparticles.  In figure~\ref{mass-and-moment}(b) we
display separately the quasiparticle mass for all iron $3d$ orbitals in the PM state
and we normalize it to its band value.  Clearly there is some
correlation between mass enhancement in Fig.\ref{mass-and-moment}(b)
and ordered magnetic moment in Fig.\ref{mass-and-moment}(a) across
various families of iron-based compounds. In particular, correlations
are too weak for ordering in LaFePO, while very heavy quasiparticles
in FeTe produce large moment of $2.1\,\mu_B$. However, there are other
factors presented below, such as kinetic frustration, orbital
differentiation and Fermi surface
shape, which together conspire to produce the magnetic orderings
displayed in Fig.\ref{mass-and-moment}(a).

The quasiparticle mass displayed in Fig.~\ref{mass-and-moment}(b) is
quite moderate in phosphorus 1111 compound on the right hand side of
Fig.\ref{mass-and-moment}(b), but correlations are significantly
enhanced in arsenic 122 and 1111 compounds. Notice however, that
enhancement is not equal in all orbitals, but it is significantly
stronger in the $t2g$ orbitals, i.e.,  $xz$, $yz$, and $xy$.  The
correlations get even stronger in 111 compounds, such as LiFeAs and
NaFeAs, and finally jump to significantly larger values of the order
of five in selenides KFe$_2$Se$_2$ and CsFe$_2$Se$_2$.
Finally, the mass enhancement of the $xy$ orbital in FeTe exceeds factor
of seven compared to the band mass, which is typical for heavy fermion
materials, but is rarely found in transition metal compounds.
We displayed only a lower bound for this mass as the end point of an
arrow in Fig.~\ref{mass-and-moment}(b), because the
quasiparticles are not yet well formed at studied temperature
T=116$\,$K.
Notice the strong orbital differentiation in FeTe, with $xz/yz$ mass
of five and $eg$ mass enhancement of only three.  This orbital
differentiation signals that the material is in the vicinity of an
orbital selective Mott transition, as proposed previously for other
iron pnictides~\cite{Medici}, where $xy$ orbital is effectively
insulating while other orbitals remain metallic. In
Fig.\ref{mass-and-moment}(b) we also display mass enhancement
extracted from optics\cite{optics-NaFeAs, optics-Ba122-Sr122,
  optics-LaFeAsO, optics-LaFePO} and ARPES\cite{PES-FeTe, PES-FeSe,
  PES-LiFeAs, PES-Ba122-Sr122, PES-Ba122, PES-Ca122} measurements, and
notice a good agreement between our theory and experiment when
available.  The effective mass extracted from ARPES and optics should
be compared with that of the $t2g$ orbitals which contribute most of
the spectral weight at low energy.

The large mass enhancement in Hund's metals is due to an orbital
blocking mechanism. If the Hund's coupling is very large, only the high
spin states have a finite probability in atomic histogram. The atomic
high spin ground state has maximum possible spin $S=2$, and is
orbitally a singlet, which does not allow mixing of the orbitals and
leads to orbital blocking, i.e., $\langle gs|d_\alpha^\dagger
d_\beta|gs\rangle=0$ when $\alpha\ne\beta$, where $|gs\rangle$ is the
atomic ground state in the $3d^6$ configuration, and $\alpha$ is the
iron orbital index. In the localized limit and in the absence of
crystal field effects, it is possible to derive a low energy effective
Kondo model, which has Kondo coupling for factor of $(2S+1)^2$ smaller
than a model without Hund's coupling~\cite{HundsTk}. Since the Kondo
temperature $T_K$ depends on the Kondo coupling $I_0$ exponentially
($T_K\propto \exp(-1/I_0)$), this results in enormous mass enhancement
of the order of $\exp(((2S+1)^2-1)/I_0)$ compared to the system
with negligible Hund's coupling (see also online material).

Having established why heavy quasiparticles form in iron pnictides and
chalcogenides, we can now study how the key parameters of the crystal
structure control the strength of correlations and other physical
properties, keeping the same on-site Coulomb interaction matrix.
The iron-pnictogen distance, displayed in
Fig.~\ref{nd-and-prob}(a), controls the overlap between iron and
pnictogen atom and hence makes iron electrons more localized
(itinerant) with increasing (decreasing) distance. The largest
distance is achieved in compounds with larger chalcogenide ion, such
as in FeTe, which results in very heavy quasiparticles, as seen in
Fig.~\ref{mass-and-moment}(b). 
The variation in distance alters the
overall bandwidth moderately.
The Hund's orbital blocking mechanism 
amplifies this variation.

The second key structural parameter is the tetrahedron shape, which is
parameterized in terms of pnictogen-Fe-pnictogen angle, displayed in 
Fig.~\ref{nd-and-prob}(a). This angle is equal to 109.5$^\circ$ for an  ideal
tetrahedron, and is much smaller in FeTe, where the Te ion is pushed
further away from Fe plane. The shape of the tetrahedron controls the
crystal field levels, which in turn control the orbital
occupancies. We display them in Fig.~\ref{nd-and-prob}(b).
The average occupation of iron atom is around
$n_d=6.25$ across all the compounds studied, which leads to an
average orbital occupation of $n_\alpha=1.25$. A deviation from ideal
angle enhances the crystal field splittings between $xy$ and
the degenerate $xz/yz$ orbital and also changes the splitting
between $t2g$ and $eg$ orbitals. Heavier quasiparticles with smaller
quasiparticle bandwidth are more susceptible to the crystal field
splitting, hence the orbital differentiation is largest in FeTe but
very small in LaFePO. The net result of crystal field splittings and
quasiparticle mass is the charge transfer from the $t2g$ to $eg$
orbitals as seen in Fig.~\ref{nd-and-prob}(b), and among $t2g$'s the
$xy$ orbital loses most charge with increased correlation strength,
pushing its occupancy closer to integer filling.

Furthermore, the effective hopping between neighboring iron atoms has
two contributions, one is due to direct iron-iron overlap, and the
second is indirect hop through pnictogen atom. The two contributions
to the diagonal hopping $t_{\alpha,\alpha}$ have \textit{opposite}
sign and destructively interfere. For the $xz$ and $yz$ orbital, the
indirect hopping through pnictogen is larger than direct iron-iron
hop. For the $xy$ orbital, the two contributions are very similar, and
when the pnictogen height is sufficiently large, such as in FeTe, the
indirect hop is reduced and the two contributions almost exactly
cancel each other, resulting in vanishing effective nearest neighbor iron-iron
$t_{xy,xy}$ hopping. This kinetic frustration mechanism contributes to
the dramatic enhancement of the $xy$ mass in the FeTe compound.

\begin{figure}[htb]
\includegraphics[width=0.99\linewidth]{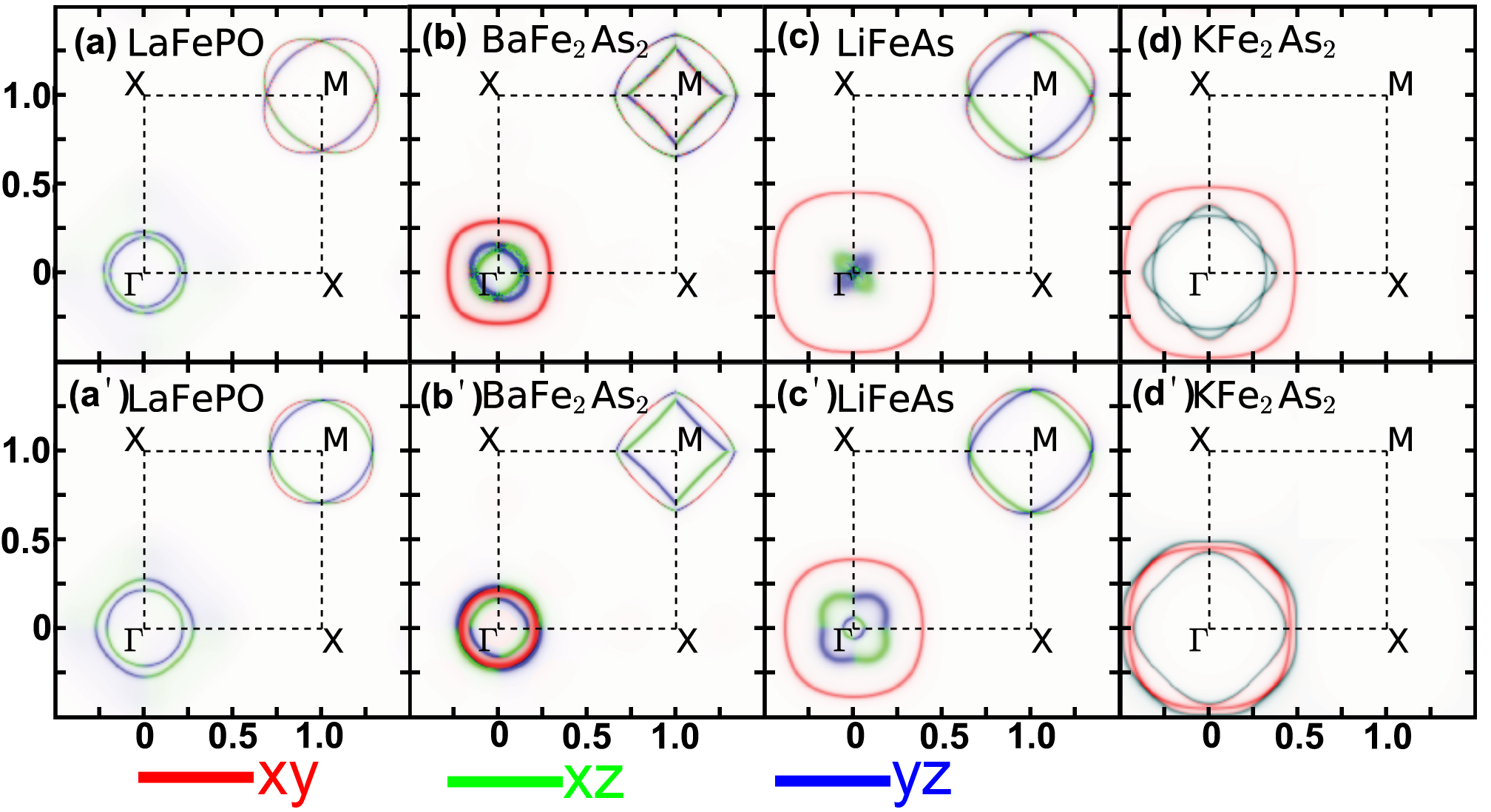}
\caption{ \textbf{Fermi surface.}  The DFT+DMFT (top row) and DFT
  (bottom row) calculated 2D Fermi surface in the $\Gamma$ plane
  ($k_z=0$) for (a)LaFePO; (b)BaFe$_2$As$_2$; (c)LiFeAs;
  (d)KFe$_2$As$_2$.
The Fermi surface is colored in red, green and blue
according to its orbital character of $xy$, $xz$ and $yz$,
respectively.}
\label{FS}
\end{figure}

In itinerant systems, the shape of the
Fermi surface, or the Fermi surface nesting is relevant for deciding
which magnetic ordering wave vector is realized when the residual
interactions among the quasiparticles is sufficiently
strong. Additional terms, arising from the incoherent part of the
electron become increasingly important as the localization threshold
is approached.

In Fig.~\ref{FS} we display DFT+DMFT Fermi
surface together with DFT predictions. In the moderately correlated end,
such as in the phosphorus 1111 compounds, our theoretical predictions
match DFT results. However, when correlations become sizable, such as
in LaFeAsO or BaFe$_2$As$_2$, the $xy$ orbital starts to play a
special role, which results in slightly modified Fermi surface shape and
character compared with DFT, while respecting the Luttinger theorem.
In BaFe$_2$As$_2$ DFT predicts that the outer pocket at $\Gamma$
is of $xz/yz$ character, while DFT+DMFT predicts that the outer pocket
is of $xy$ character, in agreement with experiments~\cite{Zhang}. This
effect of growing $xy$ pocket at $\Gamma$ and consequently shrinking
of $xz/yz$ pocket is even more apparent in LiFeAs. In the latter
compound DFT also predicts the outer pocket to be of $xy$ character,
but its size is considerably smaller than measured in ARPES
experiment~\cite{PES-LiFeAs}. DFT+DMFT increased size of the outer
$xy$ pocket together with the butterfly shape of the $xz/yz$ pocket at
$\Gamma$ are in good agreement with experiment of
Ref.\onlinecite{PES-LiFeAs}.  These changes in the shape of the Fermi
surfaces are the momentum space counterpart of the real space picture
of charge transfer among the iron $3d$ orbitals shown in
Fig.~\ref{nd-and-prob}(b). This is because the decrease (increase) of
the $xy$ ($xz$,$yz$) orbital occupancy results in the increase
(decrease) of the hole pocket size.
Finally, the Fermi surface of KFe$_2$As$_2$ displayed in Fig.~\ref{FS}
has only hole pockets around $\Gamma$ but no electron pockets at
$M$, hence there is no Fermi surface nesting to facilitate the long
range magnetic order. Indeed KFe$_2$As$_2$ can not sustain SDW
ordering and only a tiny DSDW moment can be stabilized,
as shown in Fig.~\ref{mass-and-moment}(a).
Even though the mass enhancement in KFe$_2$As$_2$ and 111
compounds is substantial, the Fermi surface nesting still plays an
important role in stabilizing magnetic ordering.

The fluctuating moment
presented in Fig.~\ref{mass-and-moment}(a) monotonically increases
with increased correlation strength, and constitutes an upper bound to
the size of the ordered magnetic moment. However, even when the Fermi
surface nesting is quite good, such as in 1111 and many 122 compounds,
the ordered moment is substantially reduced from this upper bound. A
part of the reduction is due to kinetic frustration, discussed
above. This effect is properly treated by DFT method, nevertheless the
ordered moment predicted by DFT is around 2.0 $\mu_B$ (see also online material), substantially above
the experimentally measured values. In the DFT+DMFT theoretical method,
the orbital differentiation is responsible for large overall reduction
of the static moment. In very itinerant system, such as LaFePO, the
quasiparticles are too weakly interacting to condense, hence moderate
correlations with mass enhancement of $1.5$ do not sufficiently
localize electrons to allow magnetic ordering.  In most of other
compounds, the localization and hence the effective mass increase is
substantial only in $t2g$ orbitals, while $eg$ orbitals remain only
moderately correlated. Consequently the ordered magnetic moment is
small in $eg$ orbitals, which causes only a fraction of the
fluctuating moment to order. Only when correlations are very strong
and equal in all orbitals, almost the entire fluctuating moment
orders. Such an example is provided by K$_{0.8}$Fe$_{1.6}$Se$_2$,
which is obtained by introducing iron vacancies into KFe$_2$Se$_2$,
where the entire fluctuating moment of 3.3$\mu_B$ orders.~\cite{Yin-KFeSe}

We conclude with some
experimental consequences of the theory.  We established that in
compounds with substantial mass enhancement, the $xy$ orbital is the
heaviest and most incoherent,
placing FeTe at the verge of an orbitally selective Mott
transition. Furthermore we have identified the chemical handle,
kinetic frustration, responsible for this effect. This idea can be
tested by applying uniaxial pressure on the FeTe, which should result
in a noticeable restoration of coherence in the transport properties.
Our results also suggest a natural origin for a \textit{particle-hole
  asymmetry} in doping the parent compounds.  Reducing the iron
occupancy of the $3d$ orbital, brings the occupancy of the $xy$ orbital
closer to unity, and increases the correlation strength, which in turn
strengthens the magnetic moment.  This has been observed in ARPES
studies of the BaFe$_2$As$_2$ family\cite{Yi}.

In the magnetic state, the in-plane resistivity is very anisotropic,
as has been shown in optical and transport studies of
BaFe$_2$As$_2$~\cite{Dusza, Yin-np}. This is a consequence of the
strong low energy orbital polarization of the $xz$ and $yz$
orbital\cite{Yin-np}.  Increased correlation strength results in an
increasing participation of the $xy$ orbital, with 
magnetic moment concentrated in the $xy$ orbital, which does not cause
the in-plane polarization. Hence, in-plane transport and optical anisotropy
should be reduced with increase of correlation strength and orbital
differentiation.

Our work suggests that larger mass enhancement of the $xy$ orbital
leads to smaller superconducting gap on 
the most outside hole pocket centered at $\Gamma$, 
which is mostly of $xy$ character. 
Furthermore, large degeneracy is a fertile ground for superconductivity, 
while large orbital differentiation is harmful, 
which suggests that superconductivity is hard to achieve when $X$-Fe-$X$ angle is small. 
On the other hand, large $X$-Fe-$X$ angle 
in iron pnictides and chalcogenides is accompanied by small $X$-Fe distance, 
which weakens magnetism and hence likely undermines superconducting pairing strength. 
We thus suggest that good candidates for high
temperature superconductivity are compounds with $X$-Fe-$X$ angle close to ideal
angle, as observed by Lee {\it et al.},\cite{Lee} 
in order to achieve small 
orbital differentiation, but with the largest possible $X$-Fe 
distance to strengthen spin fluctuations.

\textbf{METHOD}\\
We use fixed Coulomb interaction parameters for all materials in our DFT+DMFT calculations 
in order to keep parameter-free spirit 
and to demonstrate that the variations in the calculated physical properties 
is mainly due to the variations in the key structural parameters, rather than changes in
the screening of the Coulomb interactions. 
The detail of the method is included in online material.

\textbf{Acknowledgments} ZPY and GK were supported by NSF DMR-0906943, KH was supported by NSF DMR-0746395.
Part of the work (ZPY) was carried out under the auspices of a
DoD National Security Science and Engineering Faculty
Fellowship, via AFOSR grant FA 9550-10-1-0191.
Acknowledgment (KH) is made to the donors of
the American Chemical Society Petroleum Research Fund
for partial support of this research.

\clearpage

\setcounter{figure}{0}
\makeatletter
\renewcommand{\thefigure}{S\@arabic\c@figure}
\makeatother

\makeatletter
\renewcommand{\@biblabel}[1]{[S#1]}
\makeatother

\onecolumngrid
{\centering 
\textbf{Kinetic frustration and the nature of the magnetic and paramagnetic states in
  iron pnictides and iron chalcogenides: Supplementary online material}\\
}

\vskip 5mm

{\centering
Z. P. Yin$^{1,2}$,
\email{yinzping@physics.rutgers.edu}
K. Haule$^1$,
and G. Kotliar$^1$\\
}

{\centering
\textit{
$^1$Department of Physics and Astronomy, Rutgers University, Piscataway, New Jersey 08854, United States.\\
$^2$Department of Physics and Astronomy, Stony Brook University, Stony Brook, New York 11794, United States.\\
}
}

\vskip 5mm

\twocolumngrid
\subsection{Method}

We use the combination of density functional theory and dynamical mean
field theory (DFT+DMFT)~[S\onlinecite{DMFT-RMP2006}] as implemented in
Ref.~S\onlinecite{Haule-DMFT}, which is based on the full-potential
linear augmented plane wave method implemented in Wien2K~[S\onlinecite{wien2k}], to carry out
our first principles calculations.  The electronic charge is computed
self-consistently on DFT+DMFT density matrix. The quantum impurity
problem is solved by the continuous time quantum Monte Carlo
method~[S\onlinecite{Haule-QMC},~S\onlinecite{Werner}], using Slater form of the Coulomb repulsion in
its fully rotational invariant form.

We use the experimentally determined lattice structures, including the
internal positions of the atoms, from 
Refs.~[S\onlinecite{a-FeTe}
-~S\onlinecite{a-LaFePO}].
We use the paramagnetic tetragonal lattice structures, and neglect the
weak structural distortions. This distortion has a very small effect
on the size of the magnetic moment, proving that the magnetism has
electronic rather than structural origin.

We studied the paramagnetic phase of all compounds at the same
temperature of $T=116\,$K, and magnetic states (SDW and DSDW) at $T=72.5\,$K.  
Our \textit{ab initio} estimation for the Coulomb interaction $U$ and Hund's
coupling $J$ in BaFe$_2$As$_2$ are $U=5.0\,$eV and
$J=0.8\,$eV~[S\onlinecite{Kutepov}]. We checked that the Hund's coupling, which
is very weakly screened in solids, is very similar in other
compounds, such as FeTe, where it increases for less than 5\% compared
to BaFe$_2$As$_2$. As shown in Ref.~[S\onlinecite{Yin-np1}], physical
properties are not sensitive to small variation of Hubbard interaction
$U$, hence we fixed Coulomb interaction $U$ and $J$ to the same
\textit{ab initio} determined values ($U=5.0~$eV and $J=0.8~$eV) across all
studied compounds in the paramagnetic phase and DSDW phase, 
whereas a fixed $U=5.0~$eV and a smaller fixed $J=0.7~$eV are 
used in the SDW state in consistent with our previous calculations 
for BaFe$_2$As$_2$ in the SDW phase.[S\onlinecite{Yin-np1}]

\subsection{Histograms and Density of States}

\begin{figure}[bht]
\includegraphics[width=0.95\linewidth]{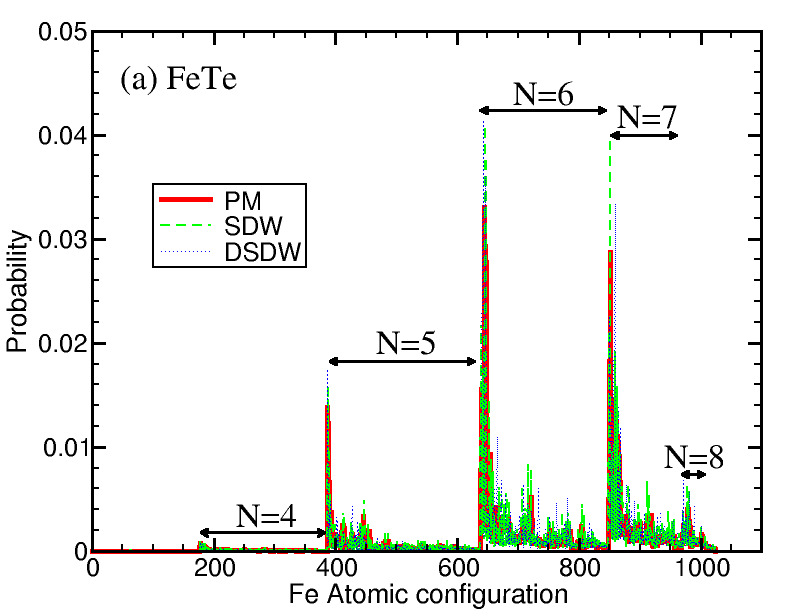}
\includegraphics[width=0.95\linewidth]{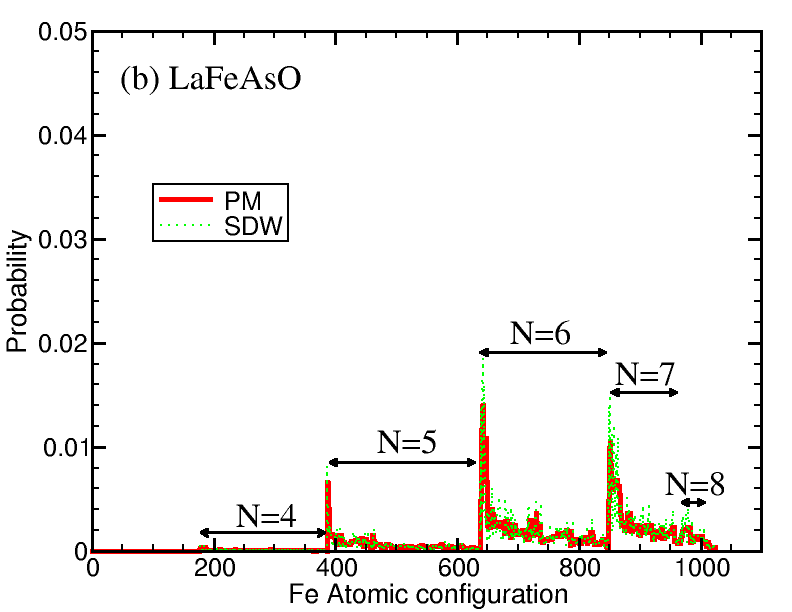}
\caption{
\textbf{Atomic histogram} 
The atomic histogram of the Fe-$3d$ shell for 
(a) FeTe and 
(b) LaFeAsO
in the paramagnetic state and magnetic states.
The 1024 possible atomic configurations are sorted 
by the number of $3d$ electrons of the individual configuration.
}
\label{histogram}
\end{figure}

To analyze the character of the many body wave function, it is
instructive to project it to momentum and real space basis. The one
electron spectral function contains the information of the overlap
between the many body wave function and the plane wave basis. It is
also instructive to project the wave function to the atomic basis on
the iron site. This projection can be presented in the form of the
atomic histogram. In any given period of time, an iron atom visits
many states from the atomic basis on the timescale, which is
proportional to the quasiparticle mass enhancement. The probability to
find an iron atom in the solid in one of the atomic states, is called
the atomic histogram, and a typical example is provided in
Fig.~\ref{histogram}. Here the atomic basis is constructed from the
five $3d$ orbitals of an iron atom, which together with spin, span a
Hilbert space of size $2^{10}=1024$. We sort these states first
according to their occupancy $N=0,1,...10$, and within the same
occupancy, we sort them according to their atomic energy. Due to large
Hund's coupling the first (last) few states at given $N$ are the high
(low) spin states. In Fig.~\ref{histogram} we clearly see the spikes
in probability for the high spin states (at the beginning of the
constant $N$ interval). Consequently, the low spin states (at the end
of the constant $N$ interval) lose substantial weight. In the absence
of Hund's coupling, the high and the low spin states would be equally
probable.  We plot in Fig.~\ref{histogram} two representative
histograms, for FeTe and LaFeAsO compounds. The two histograms are
qualitatively similar, nevertheless the differentiation between the
high-spin states and the low spin states in FeTe is more amplified.
In the magnetic states such as SDW and DSDW states, the high spin
atomic states gain even more weight, as seen in Fig.~\ref{histogram}.

\begin{figure}[bht]
\includegraphics[width=0.95\linewidth]{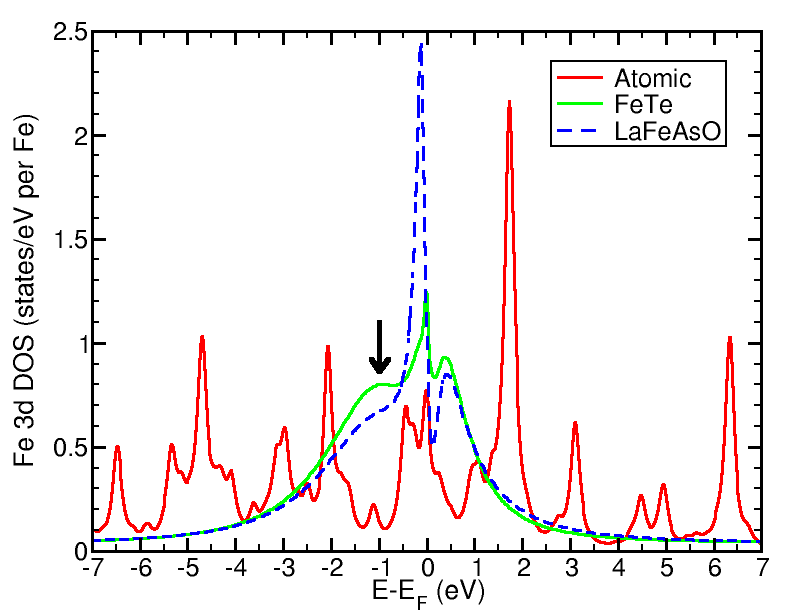}
\caption{
\textbf{Fe $3d$ DOS}
Atomic-like Fe $3d$ DOS for FeTe contrasted with actual Fe $3d$ DOS of LaFeAsO
and FeTe computed by DFT+DMFT.
}
\label{hubbard}
\end{figure}

The valence histogram of a Hund's metal is fundamentally different from
that of an oxide. While only a few atomic states have a significant
probability in an oxide, Hund's metals visit a large number of atomic
states over time, resulting in a dramatic (40\%) reduction of the
magnetic moment due to valence fluctuations. A monovalent
histogram with only the atomic ground state would give iron
magnetic moment of 4 $\mu_B$. 

Another interesting feature of Hund's metals is that very large number
of atomic states has finite probability. For comparison, in transition
metal oxides
or in heavy fermion materials with similar mass
enhancement as in iron pnictides and chalcogenides, the atomic
histogram would contain only a small number of states with significant
probability~[S\onlinecite{Shim}]. Since the Hund's rule coupling $J$ is equal to 0.8$\,$eV,
the energy spread of atomic states at constant $N=5$ or $N=6$ is
very large,
of the order of $6-7\,$eV. Because there are many
atomic states with finite probability that contribute to the one
electron spectral function, and because those states are extended
over a wide energy range, the spectral function does not have a very
well defined atomic like excitations.  To demonstrate this effect, we
plot in Fig.~\ref{hubbard}(a) an atomic spectral function of Fe $3d$ orbitals, obtained
from the corresponding atomic Green's function defined by
\begin{equation}
G(\omega) = \sum_{\alpha,m,n}\frac{|\langle n| d_\alpha^\dagger|m \rangle|^2(P_n+P_m)}{\omega-E_n+E_m}
\end{equation}
where $n$, $m$ run over all atomic states, and $\alpha$ runs over Fe $3d$
orbitals, and $P_n$ are atomic probabilities displayed in
Fig.~\ref{histogram}. Clearly, the atomic spectral weight is
distributed over a very large energy range. For comparison, a typical
heavy fermion would have one sharp peak (a delta function)
below the Fermi level, and another peak above the Fermi level, i.e., a
lower and an upper Hubbard band.[S\onlinecite{Shim}]

In Fig.~\ref{hubbard}(a) we also show the full DFT+DMFT spectral function
of the iron atom in the solid for FeTe and LaFeAsO. One can notice
that these spectral functions have a sharp quasiparticle peak close to
the Fermi level. Due to larger mass enhancement in FeTe, the
quasiparticle peak in this compound is substantially smaller than in
LaFeAsO. The rest of the spectral weight does not have a well defined
Hubbard like bands, not because the rest of the spectra would be 
coherent, but because of the unusual atomic histograms of the Hund's
metals. A small feature around $-2$ to $-1\,$eV is however noticeably
enhanced in FeTe compared to LaFeAsO. This peak was 
identified in Ref.~S\onlinecite{Ainchorn-FeSe} as an atomic-like
excitation,
which is found in atomic spectral function at $-2.2~eV$, and is related
to the excitation from atomic ground state of $d^6$ to atomic ground
state of $d^5$.

\begin{figure}[bht]
\includegraphics[width=0.95\linewidth]{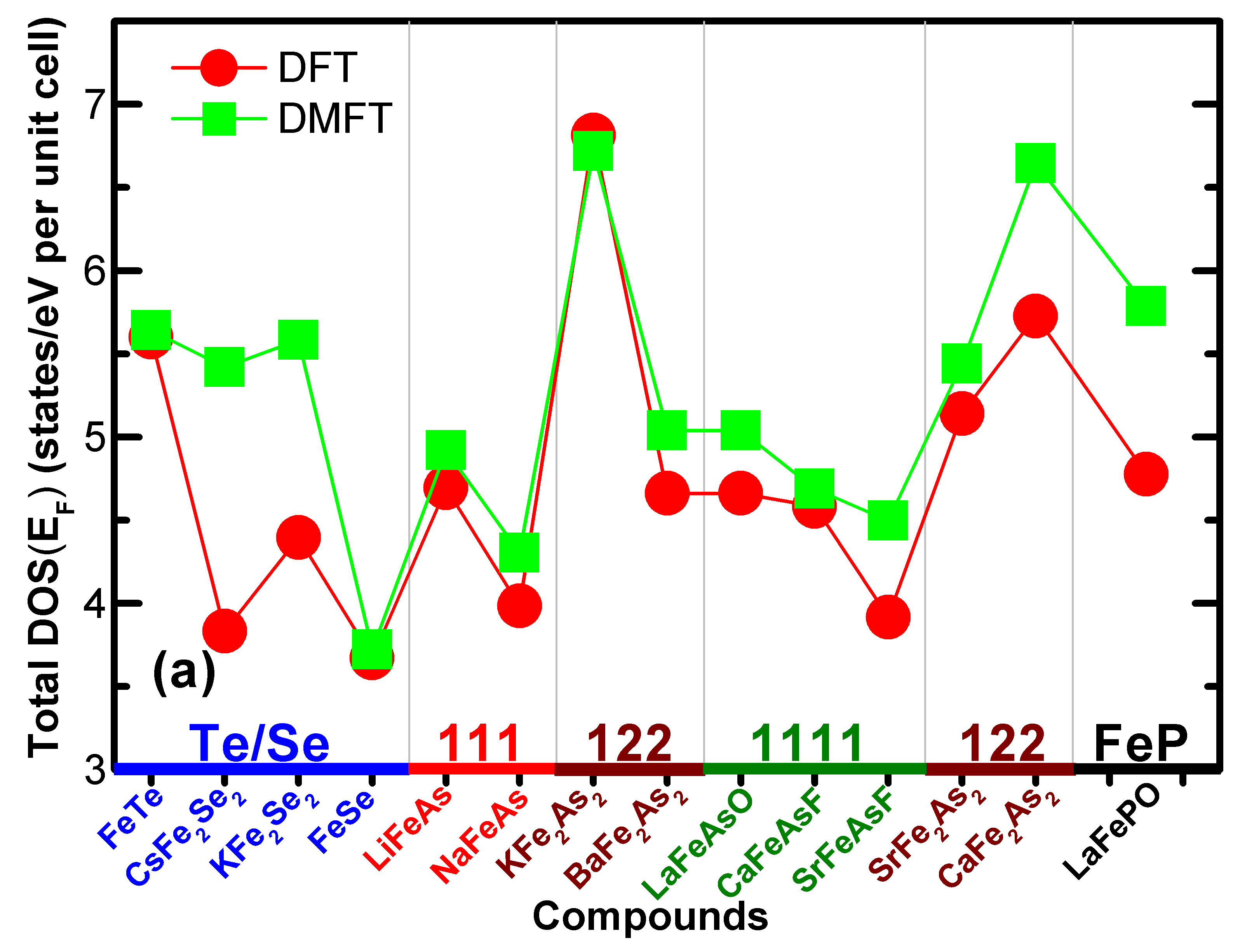}
\includegraphics[width=0.95\linewidth]{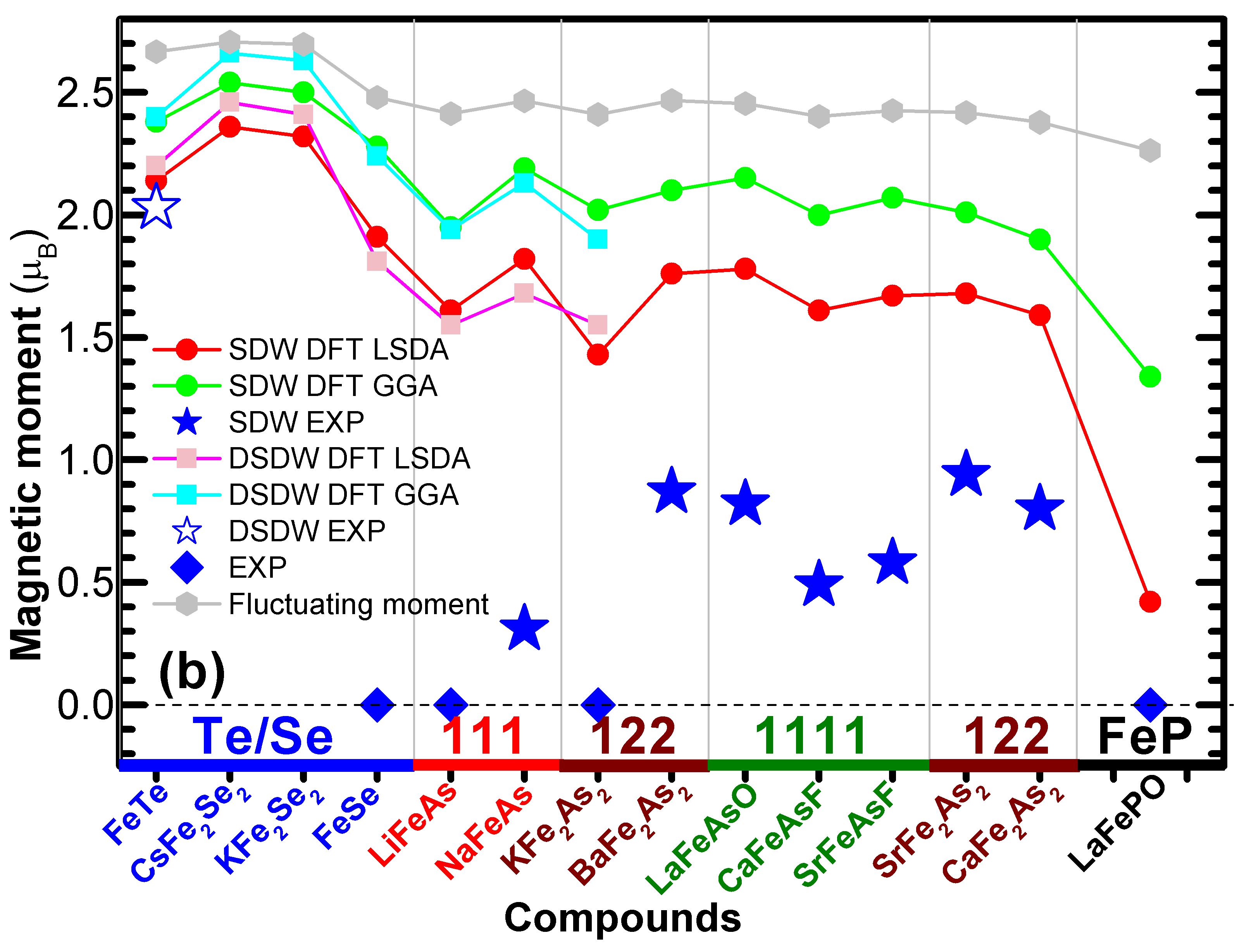}
\caption{
\textbf{DOS and magnetic moment}:
(a) Total density of states at the Fermi level in the PM phase computed by DFT and DFT+DMFT.  
(b) The magnetic moment calculated by DFT with both LSDA and GGA exchange-correlation functionals in both the SDW phase and DSDW phase. 
The fluctuating moment in the PM phase calculated by DFT+DMFT and the experimental magnetic moment in the magnetic states 
which are shown in Fig1(a) in the manuscript and reproduced here for easier comparison.  
}
\label{DOS}
\end{figure}

In the manuscript, we showed that one important factor in determining
the size of the magnetic moment is the quasiparticle mass
enhancement. Clearly the heavier quasiparticles with smaller
quasiparticle effective width are more prone to ordering. It is
interesting to inspect also the "quasiparticle height", i.e., the
value of the one-electron spectral function at the Fermi level.  In
Stoner theory, this value plays a crucial role in determining the
critical temperature and the size of the ordered moment. In
Fig.~\ref{DOS}(a) we show the value of the density of states at the
Fermi level in the paramagnetic state as obtained by both DFT and
DFT+DMFT. Clearly, large density of states at the Fermi level is more
compatible with the small moment rather than large moment (shown in
Fig.~\ref{DOS}(b)), which disfavors Stoner theory for explanation of
the trends in magnetic states across iron pnictides and chalcogenides.

We also show in Fig.~\ref{DOS}(b) the magnetic moment in the SDW and
DSDW phases calculated by DFT with both the local spin density
approximation (LSDA[S\onlinecite{PW91}]) and generalized gradient
approximation (GGA[S\onlinecite{PBE}]) exchange correlation
functionals.  We also repeat the paramagnetic fluctuating moment and
the experimental static ordered moments from the manuscript for better
comparison.  It is clear from Fig.~\ref{DOS}(b) that the DFT calculated
magnetic moments roughly follows the trend of the fluctuating moment in
the PM state, but is very different from the static ordered moment, as
already pointed out by Ref.~S\onlinecite{Held}.

\subsection{Optical properties}

\begin{figure}[htb]
\includegraphics[width=0.95\linewidth]{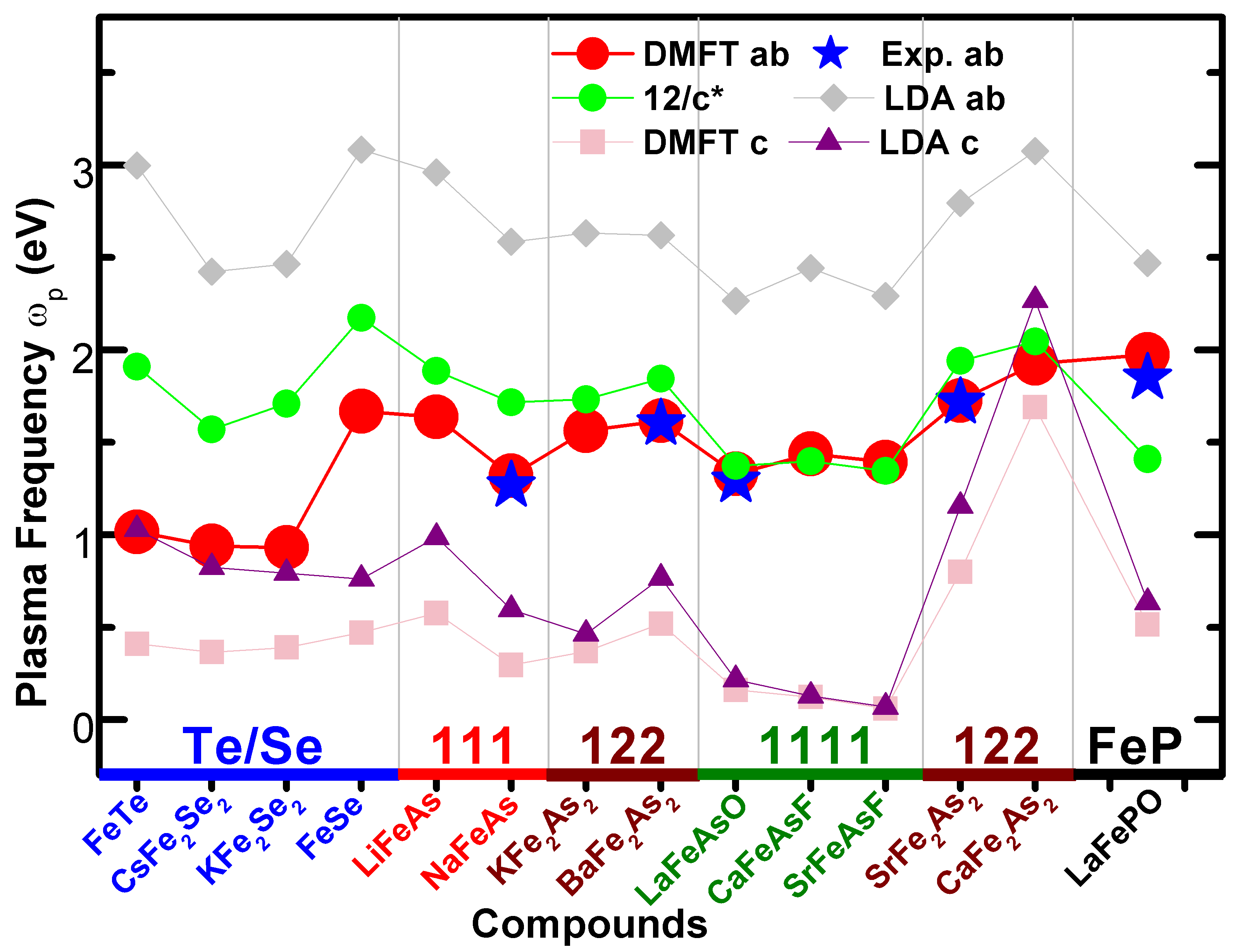}
\caption{
\textbf{Plasma frequency.}
The PM in-plane plasma frequency $\omega_{ab}$ and out-of-plane 
plasma frequency $\omega_{c}$ for various iron pnictides and iron chalcogenides calculated by both DFT+DMFT and DFT. 
The experimental PM in-plane plasma frequencies 
are taken from Refs. S\onlinecite{optics-NaFeAs1}
-~S\onlinecite{optics-LaFePO-1}. 
}
\label{plasma}
\end{figure}

Now we turn to the plasma frequencies in the paramagnetic state of
iron pnictide and chalchogenide compounds, shown in
Fig.~\ref{plasma}. We show separately the in-plane and c-axis values,
as obtained by both the DFT+DMFT and DFT calculations. We also plot the
experimentally determined in-plane values from
Refs. [S\onlinecite{optics-NaFeAs1}] for Na$_{1-\delta}$FeAs,
[S\onlinecite{optics-Ba122-Sr122-1}] for BaFe$_2$As$_2$ and
SrFe$_2$As$_2$, [S\onlinecite{optics-LaFeAsO-1}] for LaFeAsO, and
[S\onlinecite{optics-LaFePO-1}] for LaFePO.  The DFT+DMFT calculated
in-plane plasma frequencies agree well with existing optical
measurements, but are significantly reduced from the DFT values,
showing the important of correlation effect.  The extracted plasma
frequencies in the DFT+DMFT calculation for FeTe are most strongly
reduced from DFT values, and bear bigger error bars due to the fact
that the scattering rate in FeTe is so large that there is no well
defined Drude peak in the optical conductivity.
The $c$-axis Drude response is not very different from the DFT
prediction, which can be rationalized by the fact that mass
enhancement for the $z^2$ orbital is the smallest.

In Fig.~\ref{plasma} we also plot the inverse of the interlayer
distance $c^*$ of the Fe planes, multiplied by a constant factor,
i.e., $12/c^*$. There is clearly a correlation between the inverse of
the interlayer distance and in-plane Drude peak strength, suggesting
that crystal structure again plays the key role in optical response.

\subsection{Orbital blocking mechanism}

As explained in the manuscript, the orbital blocking mechanism
dramatically increases the strength of correlations in some
multiorbital systems with large Hund's coupling, when the atomic
ground state has large spin $S$.

To understand this mechanism in terms of mathematical equations, it is
useful to translate the multiorbital system to the language of a self
consistent Kondo effect~[S\onlinecite{Moeller}], describing the formation
of composite quasiparticles via a Kondo Hamiltonian having a form
$\sum_{k,k'\alpha, \beta } I_{\alpha, \beta} d_{\alpha\sigma}^\dagger
d_{\beta \sigma'} {c^\dagger}_{k'\beta, \sigma'} {c}_{k \alpha,
  \sigma} $ plus the band Hamiltonian of free electrons and the
impurity Hund's term.  The Kondo interaction is given by $I_{\alpha,
  \beta}= \langle |d_\alpha^\dagger {1 \over H} d_\beta|\rangle
+c.c. $ with $H$ an effective atomic Hamiltonian and $< >$ an average
of the most probable atomic configurations~[S\onlinecite{Moeller}]. In the
limit of no Hund's coupling, the Kondo coupling $I$ is independent of
orbital indices and the characteristic coherence scale is given by
${T_K} \propto \exp(-{1/(I N \rho)})$ where $N=2S+1$ is the orbital
degeneracy.  In the limit of large Hund's coupling, the impurity is
represented by a maximal spin $S$, and the $I_{\alpha, \beta}$ is
diagonal, since the Hund's rule coupling jams electrons into the same
spin and different orbital one particle states in the low energy many
body state.  As a result, the Kondo energy is dramatically reduced to
${T_K} \propto \exp(-{N/(I \rho)})$~[S\onlinecite{HundsTk-1}]. We clearly see
that the Hund's rule coupling has a strong effect in reducing the
coherence scale.

The crystal field splittings regulate the orbital selectivity of the
mass enhancement via the orbital selective blocking mechanism.
The more diagonal the effective Kondo coupling, the stronger the
correlations.  There are two types of high spin states in atomic
$3d^6$ configurations, namely $eg^3 t2g^3$ and $eg^2 t2g^4$. When the
low energy atomic configuration is $eg^3 t2g^3$ ($eg^2 t2g^4$), the
$t2g$ ($eg$) orbitals remain blocked, while $eg$ ($t2g$) orbitals can
mix. From this consideration it is clear that the crystal field
environment which puts $eg$ orbitals below $t2g$ makes the $eg^3
t2g^3$ primary configuration, and effectively blocks the $t2g$
orbitals, causing substantially larger effective mass for $t2g$
orbitals.  This crystal field sequence is realized in most of
compounds we studied. The only exception is LaFePO, in which $eg$
states are slightly above $t2g$, leading to slightly stronger
correlations in $eg$ orbitals.

\subsection{Kinetic frustration}

\begin{figure}[htb]
\includegraphics[width=0.95\linewidth]{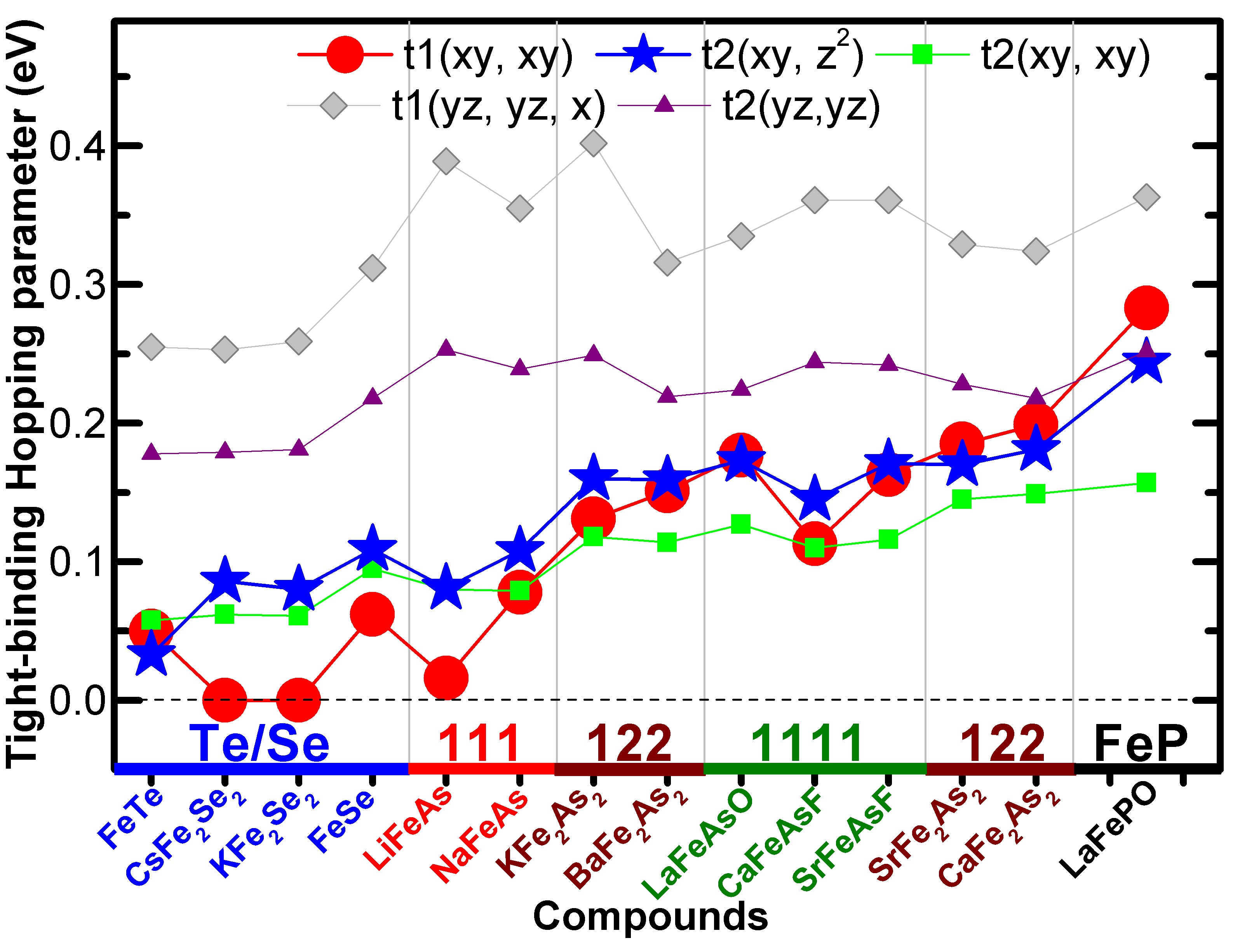}
\caption{
\textbf{Hopping parameters}
The hopping parameters of $xy$ and $yz$ orbitals to some selected orbitals of their nearest neighbor and next nearest neighbor Fe atoms obtained 
from tight-binding downfolding the DFT bands of the PM phase.
}
\label{hopping}
\end{figure}

To gain some insights into the low energy physics of iron pnictide and
chalchogenide compounds, we also constructed a low energy effective
tight-binding model. This postprocessing step is here used to
understand why a relatively small change in crystal structure can lead
to enormous difference in mass enhancement. When the model contains
both the iron $d$ electrons and pnictide/chalchogenide $p$ electrons,
the hopping parameters do not show any anomalies or specific trend
which could explain the variation of masses. On the other hand, the
effective model which contains iron $d$ electrons only, gives a clear
signature that the hopping of the electrons on the iron $xy$ orbital
is severely impeded on the more correlated end, from LiFeAs towards
FeTe. These hoppings are displayed in Fig.~\ref{hopping}.

From geometrical considerations, it is clear that direct iron
$d_{xy}-d_{xy}$ overlap between two neighboring atoms is always
negative. This is due to the sign alternation of the wave function on
the lobes of a $d_{xy}$ orbital. On the other hand, the indirect
hopping through pnictogen atom $p$ orbitals such as $p_x$, is
positive. While the indirect hopping through pnictogen $p$ orbitals is
larger than direct hopping for $xz$ and $yz$ orbitals, this
contribution is comparable for the $xy$ orbital. When pnictogen height
increases, the indirect hopping decreases, and almost exactly cancels
the direct hopping, resulting in negligible effective low energy
$xy-xy$ hopping, hence $t1(xy,xy)$ in Fig.~\ref{hopping} becomes
vanishingly small in the FeTe end. It is important to note that the bandwidth of the
$xy$ orbital does not change very dramatically, because the
off-diagonal hopping $t1(xy,xz)$ remains large (0.22-0.24$\,$eV) even
in correlated compounds. Nevertheless, the most important diagonal
hopping $t1(xy,xy)$ dramatically decrease together with the next
nearest neighbor $t2(xy,xy)$ and off diagonal next nearest neighbor
$t2(xy,z^2)$ hopping. Due to this kinetic frustration mechanism the
diagonal hopping is small in correlated compounds such as FeTe. Due to
orbital blocking mechanism, the mixing of the orbitals is blocked, and
hence the correlations can increase dramatically in the $xy$ orbital.

\end{document}